\documentclass{elsart}

\usepackage{graphicx}

\usepackage{amssymb}

\journal{Journal of Magnetism and Magnetic Materials}

\begin{document}

\begin{frontmatter}



\title{Superparamagnetic relaxation in Cu$_{x}$Fe$_{3-x}$O$_{4}$ $(x=0.5$ and $x=1)$ nanoparticles }


\author[zagreb]{Damir Paji\'{c}\corauthref{cor}}, 
\corauth[cor]{Corresponding author.} 
\ead{dpajic@phy.hr}
\ead[url]{www.phy.hr}
\author[zagreb]{Kre\v{s}o Zadro} ,
\author[gent]{Robert E. Vanderberghe}
\author[sofia]{Ivan Nedkov}
\address[zagreb]{Physics Department, Faculty of Science, University of Zagreb,
Bijeni\v{c}ka c. 32, HR-10000 Zagreb, Croatia}
\address[gent]{Department of Subatomic and Radiation Physics, University of Gent, Proeftuinstraat 86, B-9000 Gent, Belgium}
\address[sofia]{Institute of Electronics, Bulgarian Academy of Sciences,
Tzarigradsko Chaussee 72, BG-1784 Sofia, Bulgaria}

\begin{abstract}
The scope of this article is to report very detailed results of
the measurements of magnetic relaxation phenomena in the new
Cu$_{0.5}$Fe$_{2.5}$O$_{4}$ nanoparticles and known CuFe$_{2}$O$_{4}$
nanoparticles. The size of synthesized particles is (6.5$\pm $1.5)nm. Both samples show the superparamagnetic behaviour, with the well-defined phenomena of blocking of magnetic moment. This includes the splitting of zero-field-cooled and field-cooled magnetic moment curves, dynamical hysteresis, slow quasi-logarithmic relaxation of magnetic moment below blocking temperature. The scaling of the magnetic moment relaxation data at different temperatures confirms the applicability of the simple thermal relaxation model. The two copper-ferrites with similar structures show significantly different magnetic anisotropy density and other magnetic properties. Investigated systems exhibit the consistency of all obtained results. 
\end{abstract}

\end{frontmatter}


\section{Introduction}
\label{intro}

Nanosized magnetic particles have received considerable attention in
experimental, theoretical and computational solid state physics due to their rich and often unusual experimental behaviour. Assemblies of these particles are a subject of intense research for many years and the investigations are still running with the new outcomes.

Besides looking for new physics, big part of the magnetic nanoparticles
research receives impulsion from the profitable technological application of
these new and complex materials, mainly in the magnetic data-storage (memory) and biotechnology (drug delivery, self-assembly).

Today topics concerning the physics and magnetism of ultrafine magnetic particles are directed mainly to the coupling of neighbour particles
\cite{interscaling,interallcomp,interdipexch}, surface spin glass ordering
\cite{glassprl,glassprb}, influence of the mechanical treating on the properties of fine magnetic particles \cite{mechan}, memory effects \cite{memory,memoerasing}, collective phenomena \cite{collective} and very famous quantum tunneling of magnetization \cite{qtmprl,qtm94}. The aim of this paper is to compare the properties and phenomena in materials of different chemical composition but in the form of the fine particles with similar sizes and having very similar crystal structure.

The systems investigated in this work are the inverse spinel ferroxide nanoparticles of copper-ferrite with initial formula Fe$[$Fe$_{2-x}$Cu$_{x}]$O$_{4}$, where $x=0.5-1.0$. Their inverse spinel structure with two different crystallographic sublattices for the cations can be represented by a general formula A$[$B$_{2}]$O$_{4}$, where A belongs to the tetrahedral, and B to the octahedral sites. In CuFe$_{2}$O$_{4}$ the Fe$^{3+}$ cations complete the A sites and half of the B sites, and the Cu$^{2+}$ cation generally occupies the B sites \cite{japcufe2o4,nanostructcufe2o4} and is a typical Jahn-Teller ion.

The nanoparticles could be considered as a state of matter on the borderline
between the atomic and crystalline states and it is important to investigate
magnetic oxides with the similar chemical composition and low anisotropy, but with crystallographic deviations such as those occurring at the structural transition due to cooperative Jahn-Teller distortion. Particles were obtained by soft chemical process: by applying a magnetic field during the co-precipitation process, we separated the particles with spherical shape and particle size up to (6.5$\pm $1.5)nm. It is well known that the magnetic properties of the ferrospinels are very sensitive to the cation distribution. The compositions chosen offer the possibility to investigate the influence of the ions distribution in a spinel structure on the magnetization when the particle is spherical and single domain. Thus, we investigated the influence of the phase and structural particularities of nanosized granular inverse spinels on the magnetization processes.

Besides ferrimagnetic ordering, which originates from the strong exchange
interaction between atoms, these materials exhibit also magnetic anisotropy,
resulting from the relativistically small interaction between the atomic moments and crystal field \cite{lanlif}. Particles of actual size are preferably single domain \cite{singledomain}, so that the magnetic state of the particle can be described by one unique quantity - it's magnetic moment. Because of anisotropy, this moment has the energetically preferable orientation along the so-called easy axis. Stable orientations are separated by magnetic anisotropy barrier $U$, which ensures bistability. The relaxation of the magnetization $M$ of the ensemble of particles with the same anisotropy barriers is exponential in time. In case of the thermal relaxation, the relaxation-time $\tau $ at the temperature $T$ is given according to Ref. \cite{fluctdomain} by Arrhenius law 
\begin{equation}
\tau = \tau _{0} \e ^{U/kT},
\label{tau}
\end{equation}
where $k$ is Boltzmann constant, and $\tau _{0}\approx 10^{-9} - 10^{-11}$ is time between the attempts of escape over the barrier.

For high $T$, $\tau $ is very small compared to the time of observation of $M$,
so that $M$ fluctuates quickly over the barrier. If $T$ is low enough, $M$
remains blocked on one side of the barrier for a long time. The boundary between these two regimes depends on the time scale of the measurement.

Here, the results of very detailed magnetic measurements of substances
Cu$_{0.5}$Fe$_{2.5}$O$_{4}$ and CuFe$_{2}$O$_{4}$ in the form
of nano-sized particles are reported comparatively. They exhibit qualitatively
the same magnetic behaviour and quantitatively different properties (the blocking temperature, magnetic anisotropy and coercivity). The performed detailed measurements of the time-relaxation of the magnetic moment of the samples provides calculation of the magnetic viscosity and anisotropy, and also confirms that the relaxation follows from thermal activation.

\section{Experimental procedure and samples}
\label{exp}

The investigated Cu$_{0.5}$Fe$_{2.5}$O$_{4}$ and CuFe$_{2}$O$_{4}$ nanoparticles were prepared following basically the same procedure. The cupro-ferrite particles were formed by adding NaOH to a water solution of FeCl$_{2}\cdot$4H$_{2}$O and CuCl$_{2}$ mixed in strictly fixed concentration ratios to achieve a variation of the composition with respect to the Cu$_{x}$Fe$_{3-x}$O$_{4}$ initial formula for $x$=0.5-1.0 and co-precipitation in an alkaline medium at $pH>10$. The technological conditions were developed to apply the method of fabricating CuFe$_{2}$O$_{4}$ nano-powders with and without cooperative Jahn-Teller distortion (this was achieved through varying the cations ratio in the initial solution to 1.0), and avoid the additional annealing of the samples. The other sample object of our investigation have the composition Cu$_{0.5}$Fe$_{2.5}$O$_{4}$.

In our work, we used particles separated by means of applying a magnetic field
after the first precipitate settled down following the introduction of the
oxidant and the increase of the $pH$ of the solution. The particles thus
produced had the shape of a sphere or a prolate spheroid at sizes below 10 nm.

The X-ray phase analysis was performed by using a TUR-M62 apparatus with
Co$K\alpha $ radiation. The exact position of the lines and their widths were
determined by the computer software for deconvolution and profile analysis of
diffraction patterns. The XRD data exhibited consistently a single-phase spinel
structure for all types of samples. The TEM bright-field images and the
grain-size statistics of the powders obtained were analyzed, too.

The M\"{o}ssbauer spectra were taken with an electromechanical spectrometer
working at a constant acceleration mode from 4.5 K to room temperature. A
70mC - $^{57}$Co(Cr) source and $\alpha $-Fe  standard  were  used. The experimentally obtained spectra were computer fitted to a series of  Lorentzian curves by the least-squares method.

For the present magnetic measurements the powder of nanoparticles in molten liquid paraffin was dispersed and then cooled whilst vigorously shaking. It is
assumed that the particles in rigid paraffin stay fixed. This excludes the
mechanical rotation of the particles as a possible way of changing of their magnetic moment vector.

Magnetic measurements were performed using commercial MPMS5 SQUID magnetometer, which uses extraction through the induction coils to measure the magnetic moment of the sample. It provides high stability of temperature at the place of the sample and stable homogeneous magnetic field. Due to the SQUID detection system, magnetic moment is measured with very high accuracy. Besides the good
temperature dependence measurements, this equipment is therefore very suitable
for the slow and long-lasting relaxation measurements.

At first, magnetic measurements include the dependence of magnetic moment of the sample $m$ on the temperature $T$. The sample was cooled down to 1.8K from room temperature in zero applied field. Then the magnetic field was applied and the variation of $m$ was measured during the increasing of $T$ up to 290K. This
gave the so called zero-field-cooled (ZFC) curve. After that, the sample was
cooled again to 1.8K, but now staying at the same applied magnetic field, and
after that the variation of $m$ during the increasing of $T$ was measured. This gave the so called field-cooled (FC) curve. The temperature below which the splitting of ZFC and FC curves appears because of the slow relaxation is called blocking temperature $T_{B}$.

Secondly, $m(H)$ curves for the applied magnetic field $\mu _{0}H$ up to 5.5T at different stable temperatures were measured. Also, $m(T)$ dependence in the highest attainable applied magnetic field $\mu _{0}H$=5.5T was measured.

Lastly, very detailed and precise measurement of the relaxation of magnetic
moment of the sample at broad range of stable temperatures from
2K up to $T_{B}$ of each sample was performed. Every sample was at first heated above $T_{B}$ in zero applied magnetic field. Then the magnetic field of 0.03T was imposed. After some waiting time the sample was cooled down to the desired temperature and stabilised. Finally, the magnetic field was reversed to the opposite direction (from 0.03T to -0.03T) and $m$ was measured as the time elapsed during $\sim $3 hours. This procedure was repeated for many different target temperatures below $T_{B}$.

The measured magnetic moment $m$ of the sample is not convenient for the comparison with other published data. Therefore, the magnetization $M$ is used according to circumstances.

\section{Results}
\label{results}

The systematic study of CuFe$_{2}$O$_{4}$ nanostructured powders was laid out
in more details in our previous work \cite{supermater}. Here we summarize basic data obtained from X-ray diffraction patterns at room temperature and from the
analysis of Bragg line-broadening and grain-size statistics from high-resolution photographs of different powders. The systematic processing of the data demonstrated that the amount of particles with spherical or prolate-spheroid shape exceeded 90\%.

The objects of the mentioned studies were spherical nanosized particles of
Cu$_{x}$Fe$_{3-x}$O$_{4}$ formed via a soft chemical process. Particle sizes are in the order of $(6.5\pm 1.5)$nm. For values of $x$ of up to 0.75, we obtained cubic crystal lattice and the ratio of lattice parameters $c/a = 1$. For $x = 1$ we produced particles, whose X-ray diffraction data indicated $c/a = 1.04$, which is a sign of the existence of the Jahn-Teller distortion (for bulk Cu-ferrospinel Jahn-Teller effect usually appears at $x > 1$) and has probably been due to the Laplace pressure. We observed a distinct difference for Cu$_{0.5}$Fe$_{2.5}$O$_{4}$ with cubic crystal lattice, where $c/a = 1$.

\begin{figure}
\begin{center}
\includegraphics*[width=7.5cm]{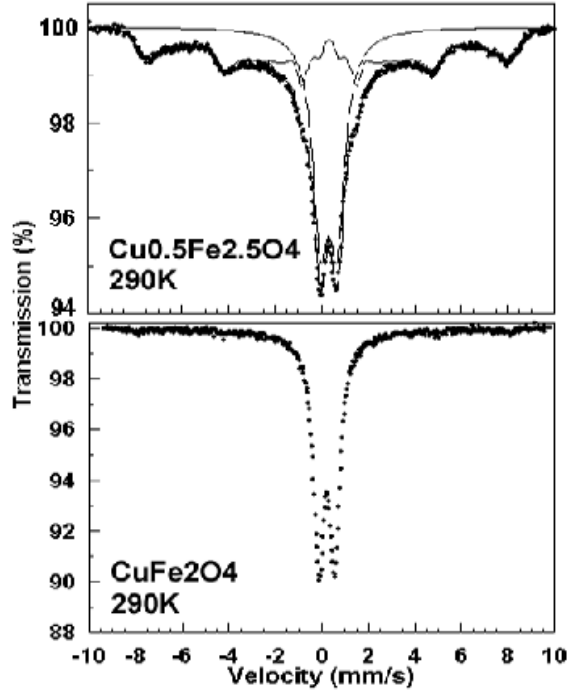}
\end{center}
\caption{The M\"{o}ssbauer spectra of Cu$_{0.5}$Fe$_{2.5}$O$_{4}$ and CuFe$_{2}$O$_{4}$ at room temperature.}
\label{moess290}
\end{figure}

Fig. \ref{moess290} shows the M\"{o}ssbauer spectra of the two samples at room  temperature that point to their clear superparamagnetic behaviour.

\begin{figure}
\begin{center}
\includegraphics*[width=7.5cm]{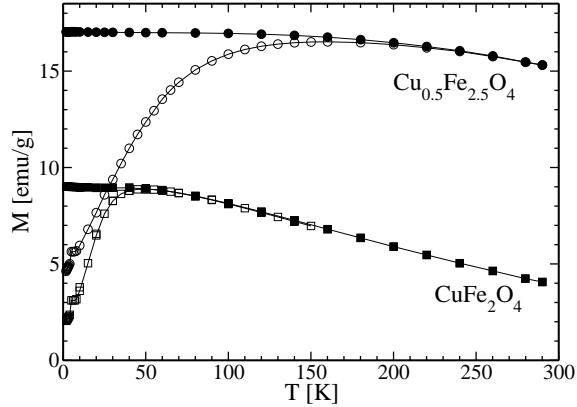}
\end{center}
\caption{The dependence of magnetization $M$ on temperature $T$ of Cu$_{0.5}$Fe$_{2.5}$O$_{4}$ and CuFe$_{2}$O$_{4}$ for the zero-field cooling mode (ZFC, open symbols) and field cooling mode (FC, full symbols) in the applied magnetic field of 0.03T. The lines are eye-guides.}
\label{zfcfc}
\end{figure}

The measurement of $m(T)$ dependence is shown in Fig. \ref{zfcfc}. The
splittings of the ZFC and FC curves are very pronounced providing the easy and
unique determination of blocking temperature $T_{B}$, which equals to 190K for Cu$_{0.5}$Fe$_{2.5}$O$_{4}$ and 50K for CuFe$_{2}$O$_{4}$. The curves are measured in applied magnetic field of 0.03T, which is small enough compared with the magnetic anisotropy field so that it does not destroy the anisotropy barrier.

\begin{figure}
\begin{center}
\includegraphics*[width=7.5cm]{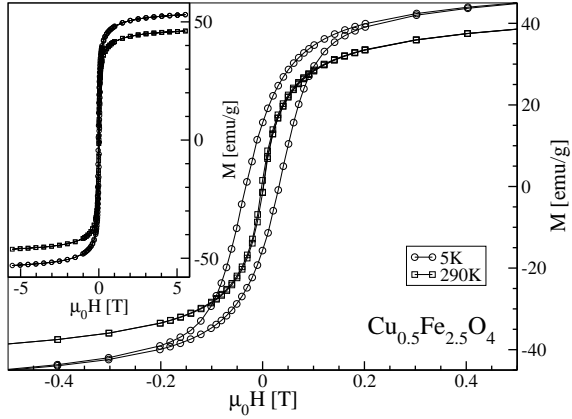}
\end{center}
\caption{Magnetic hysteresis curves $M(H)$ for Cu$_{0.5}$Fe$_{2.5}$O$_{4}$ measured at 5K and 290K. The lines are eye-guides.}
\label{hyster}
\end{figure}

Below $T_{B}$ hysteresis loops are obtained. On the contrary, above
$T_{B}$ the $m(H)$ curves are completely reversible. The $m(H)$ curves at two characteristic temperatures (5K - well below $T_{B}$, and 290K - well above $T_{B}$) for Cu$_{0.5}$Fe$_{2.5}$O$_{4}$ are shown in Fig. \ref{hyster}. Similar curves are obtained for CuFe$_{2}$O$_{4}$. The differences are just in the saturation value $M_{S}$ of magnetization, in the remanent magnetization $M_{R}$ at $H=0$, and in the value of coercive field $H_{C}$. For further reference, these data are reproduced in Table \ref{tabprop}.

\begin{figure}
\begin{center}
\includegraphics*[width=7.5cm]{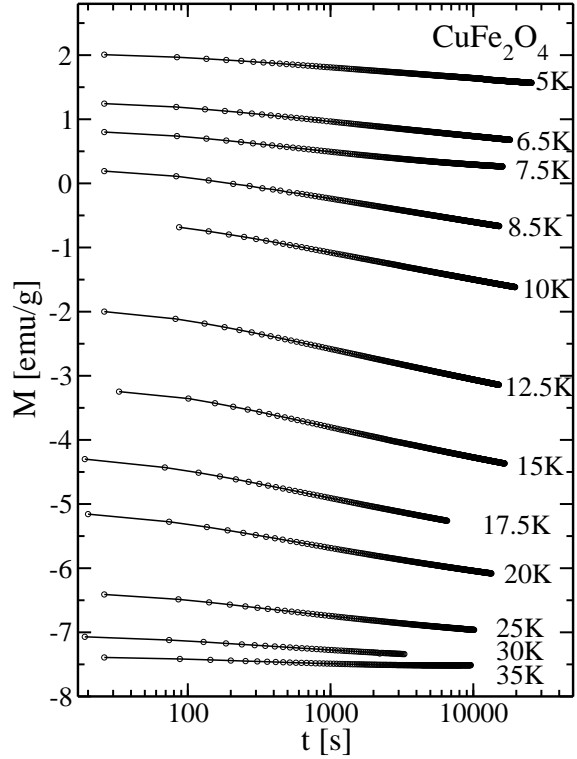}
\end{center}
\caption{Relaxation of magnetization $M$ in time $t$ for CuFe$_{2}$O$_{4}$
sample at different temperatures $T$ measured after the cooling from room-temperature in 0.03T and reversing of magnetic field to -0.03T. The lines are eye-guides.}
\label{relax}
\end{figure}

The relaxation of magnetic moment of the samples is measured below their
blocking temperatures as described in Sect. \ref{exp}. Points were sampled at
equidistant times to provide for the homogeneous conditions during the
measurement. When the dependence of magnetic moment $m$ on logarithm of time
$\ln t$ is plotted, the linear decreasing during the time of up to
$\sim $3 hours for all of the measured temperatures is observed, as shown in Fig. \ref{relax} for CuFe$_{2}$O$_{4}$. The slope of these lines, which is obtained by very reliable linear regression, is dependent on temperature. Above the blocking temperature the relaxation of magnetic moment is not observable using the magnetometer, but the constant value is measured, because $m$
already relaxed to this equilibrium value very quickly. For Cu$_{0.5}$Fe$_{2.5}$O$_{4}$ the linear dependence $m(\ln t)$ during $\sim $3 hours of relaxation measurement at each temperature is observed, too.

\begin{table}
\caption{Some magnetic property parameters of measured samples: blocking temperature $T_{B}$ in 0.03T, magnetic anisotropy
constant $K$, magnetization $M_{S1}$ at 5.5T and 5K, magnetization $M_{S2}$ at
5.5T and 290K, coercive field $H_{C}$ at 5K, remanent magnetization $M_{R}$ at 5K.}
\begin{tabular}{|c|c|c|c|c|c|c|}
\hline Sample & $T_{B}[$K$]$ & $K[$J/m$^{3}$
$]$ & $M_{s1}[$emu/g$]$ & $M_{s2}[$emu/g$]$ & $\mu _{0}H_{C}[$T$]$ & $M_{R}[$emu/g$]$ \\
\hline Cu$_{0.5}$Fe$_{2.5}$O$_{4}$ & 190 & 2.7$\cdot 10^5$ & 53.1 & 46.1 & 0.032 & 15.6 \\
\hline CuFe$_{2}$O$_{4}$ & 50 & 0.72$\cdot 10^5$ & 35.6 & 27.8 & 0.041 & 8.3 \\ 
\hline
\end{tabular}
\label{tabprop}
\end{table}

\section{Discussion}
\label{discussion}

The splitting of ZFC and FC curves is reflection of the magnetic anisotropy,
and the blocking temperature $T_{B}$ provides the estimation of the density of
magnetic anisotropy energy.

In the assembly of nanoparticles there is a distribution of energy barriers. That is confirmed in magnetic measurements of our samples, too, as described later in text. Besides the temperature dependence of the relaxation time $\tau $ for magnetic moment of the particle, there is also strong dependence of $\tau $ on the energy barrier height, as follows from Eq. \ref{tau}. Starting from lowest $T$ on the measurement of ZFC curve in field of 0.03T, $\tau $ for most of the particles is very long. Increasing $T$, $\tau $ becomes shorter, so that more and more particles relax to the equilibrium magnetization, especially those particles with lower energy barriers. At $T_{B}$ the magnetic moments of all particles relax to the equilibrium during the time of measurement of one point. Following these arguments, for relaxation time at $T_{B}$ the time $\tau =100s$ of one measurement is chosen. From many literature sources (see Ref. \cite{qtm94}) for superparamagnetic systems $\tau _{0} = 10^{-10}s$ is used, which is connected to the temperature independent fast transitions, for example like in the ferromagnetic resonance. Then follows the anisotropy barrier height $U/k=T_{B}\ln (\tau / \tau _{0})\simeq 28\cdot T_{B}$, which is 5320K for Cu$_{0.5}$Fe$_{2.5}$O$_{4}$ and 1400K for CuFe$_{2}$O$_{4}$.

To calculate the magnetic anisotropy energy density $K$, the volume of the
particle is needed, according to $U=K\cdot V$. In reality, $U$ is not simply
proportional to $V$, but $U$ depends on the shape and surface structure of the
particles. Nevertheless, this simple equation will be used to calculate the 
effective density of magnetic anisotropy energy $K$.

So, the relaxation time depends on particle's volume: for bigger particles it is longer. Also, for bigger particles it is necessary to go to the higher temperatures to have the same relaxation time as for smaller particles. Above $T_{B}$ all particles relax at least faster than one measurement, making the ZFC and FC curves to overlap. At lower temperatures the biggest
particles do not relax completely during one measurement. It was found that there exists the activation volume \cite{activvol}. The conclusion is that $T_{B}$ is defined by concerning the biggest particles in the sample. This fact is very suitable because the top diameter in the distribution of particles in our case is known: it equals to 8nm. Using the volume $V_{max}=4\pi r_{max}^{3}/3$ of spherical particles, $K=U/V_{max}$ is estimated to 2.7$\cdot 10^{5}$J/m$^{3} $ for Cu$_{0.5}$Fe$_{2.5}$O$_{4}$ and 0.72$\cdot 10^{5}$J/m$^{3} $ for CuFe$_{2}$O$_{4}$. The values calculated in this way are the effective densities of magnetic anisotropy energy, which include besides crystalline anisotropy also the shape anisotropy and more complex surface effects. It is therefore clear why this $K$ is different than the
density of magnetic anisotropy energy of bulk material, which for
CuFe$_{2}$O$_{4}$ is $6\cdot 10^{3}$J/m$^{3}$ (see Ref. \cite{jpcm}). Obtained value for the effective $K$ in CuFe$_{2}$O$_{4}$ is somewhat greater than the other published data obtained from coercive field \cite{jpcm}. The difference can come from the different methods of preparation of the samples. Also, it was found that $K$ depends on the size of particles \cite{sizeeffect}, which should be taken into account if comparing the data.

The temperature of the splitting between the FC and ZFC curves differs slightly
from the temperature at which maximum in ZFC curve appears. That is due to the
combination of the distribution over sizes of particles and the experimental
procedure. Some of the particles with lower barrier do not relax completely to
the equilibrium before the relaxation of particles with higher barriers becomes
observable, but particles with higher barriers start before than the particles
with lower barriers finish their relaxation. The difference of these two
temperatures does not affect the results remarkably.

First obvious consequence of slow relaxation is hysteresis loop below
$T_{B}$. It does not come from the domain-wall motion and their pinning as is
the case in bulk magnets, but it comes from the slow relaxation. The slow
dynamics origin of hysteresis was confirmed by measurements and modelling of 
similar systems \cite{dynamichyster}. Applied magnetic field increases the relaxation rate by lowering the magnetic anisotropy barrier. Above certain fields the $m(H)$ curve becomes reversible because the relaxation becomes faster than the measurement of one point. Completely reversible $m(H)$ curves  above $T_{B}$ are in accordance with the fast superparamagnetic fluctuations. For single size monodomain particles the magnetization follows the
Curie-Brillouin-Langevin dependence on the magnetic field.

\begin{figure}
\begin{center}
\includegraphics*[width=7.5cm]{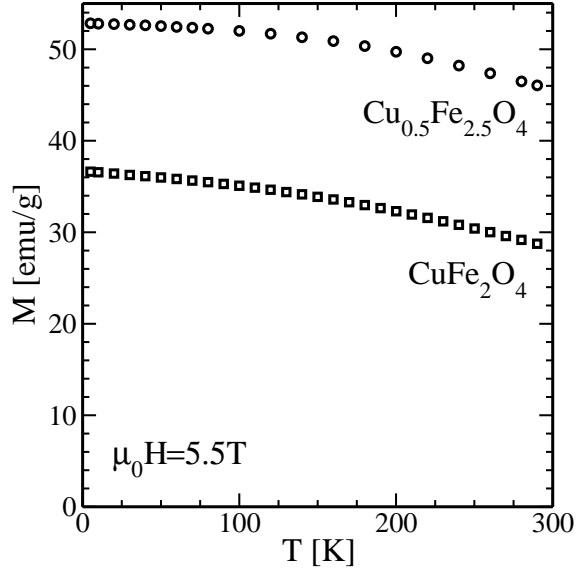}
\end{center}
\caption{The dependence of magnetization $M$ on temperature $T$ measured at applied magnetic field of 5.5T for Cu$_{0.5}$Fe$_{2.5}$O$_{4}$ and CuFe$_{2}$O$_{4}$.}
\label{mt55000G}
\end{figure}

The highest magnetic field during measurement of the hysteresis loops was 5.5T.
At that field the saturation was not attained, probably because of the spin
canting \cite{japcufe2o4}. Anyways, 5.5T is well above the point at which the
reversibility starts. It is indicative to look at the value of magnetization at
some specific magnetic field like 5.5T. For Cu$_{0.5}$Fe$_{2.5}$O$_{4}$ the magnetization decreases from 53.1emu/g at 5K to 46.1emu/g at 290K and for CuFe$_{2}$O$_{4}$ from 35.6emu/g at 5K to 27.8emu/g at 290K. The results for CuFe$_{2}$O$_{4}$ are consistent with the data published in Refs.
\cite{japcufe2o4,nanostructcufe2o4,jpcm}. The decrease of magnetization with temperature is monotonous, as shown in Fig. \ref{mt55000G}. Such a result is expected for the superparamagnetic particles.

\begin{figure}
\begin{center}
\includegraphics*[width=7.5cm]{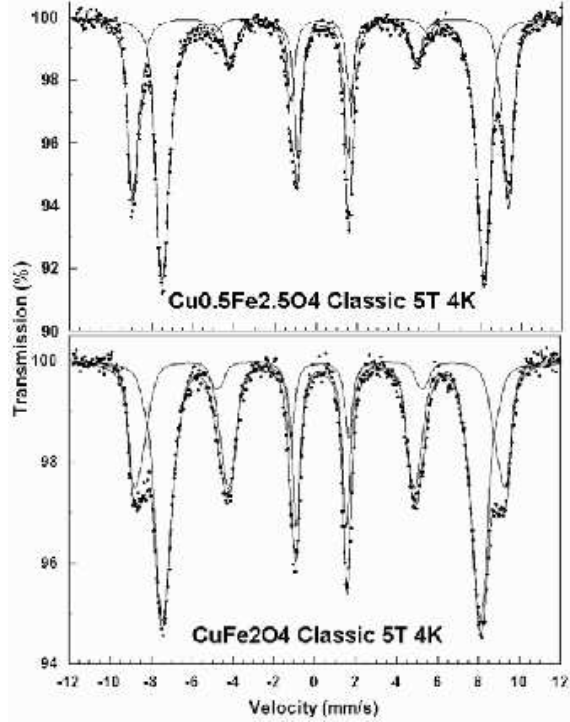}
\end{center}
\caption{The M\"{o}ssbauer spectra of Cu$_{0.5}$Fe$_{2.5}$O$_{4}$ and CuFe$_{2}$O$_{4}$ in the applied magnetic field of 5T at the temperature 4K.}
\label{moess4}
\end{figure}

The magnetization of the sample in the same applied magnetic field and at the
same temperature is different between the two samples. For the structure of the unit cell of Cu$_{0.5}$Fe$_{2.5}$O$_{4}$ two models are known \cite{cellmodel}, represented as:
\begin{equation}
\mathrm{Cu}^{1+}_{x}\mathrm{Fe}^{3+}_{1-x}[\mathrm{Fe}^{3+}_{1.5+x}
\mathrm{Cu}^{1+}_{0.5-x}]\mathrm{O}_{4} 
\end{equation}
and
\begin{equation}
\mathrm{Cu}^{2+}_{y}\mathrm{Fe}^{3+}_{1-x}[\mathrm{Cu}^{2+}_{0.5-y}
\mathrm{Fe}^{2+}_{0.5}\mathrm{Fe}^{3+}_{1+y}]\mathrm{O}_{4} .
\end{equation}
This system can exhibit a high magnetic moment (in the order of $4.5\mu _{B}$ per unit cell) at values of $x=0.2$ and $y=0.25$. Fig. \ref{moess4}  presents M\"{o}ssbauer spectroscopy data for our sample at 4K. The calculations based on these data show the following distribution of the cations in the spinel studied: 
\begin{equation}
(\mathrm{Cu}^{1+}\mathrm{Cu}^{2+})_{0.09}\mathrm{Fe}^{3+}_{0.91} [\mathrm{Fe}^{3+}_{1.59}(\mathrm{Cu}^{1+}\mathrm{Cu}^{2+})_{0.41}]
\mathrm{O}_{4} .
\end{equation}

In our case, speaking of point of view of these models, an intermediate cation distribution is realized, where one large part of the Fe$^{3+}$ cations is located in the octahedral sublattice. This is the probable explanation of the higher total saturation magnetization of Cu$_{0.5}$Fe$_{2.5}$O$_{4}$ nanostructured particles. 

Fig. \ref{moess4} presents the M\"{o}ssbauer spectrum of the composition CuFe$_{2}$O$_{4}$, too, and the cation distribution derived from it is: 
\begin{equation}
\mathrm{Cu}^{2+}_{0.43}\mathrm{Fe}^{3+}_{0.57}[\mathrm{Fe}^{3+}_{1.43}
\mathrm{Cu}^{2+}_{0.57}]\mathrm{O}_{4} .
\end{equation}
Obviously, a quasi-similar distribution of the Cu$^{2+}$ cation exists in both sublattices and the Jahn-Teller effect observed is caused by the Laplace pressure in the spherical particle.

During the time of $\sim $3 hours of the measurement at constant temperature the relaxation of magnetic moment of both samples is logarithmic. It is valid for all of the measured temperatures during this time of observation of relaxation. Therefore, it is suitable to extract the temperature dependence of the slopes of these straight lines. The logarithmic time-dependence is here a consequence of the distribution of barrier heights among the nanoparticles. Here is the short overview of this claim, whose details can be found in Ref.  \cite{mqtmm}. Considering the ensemble of the uniaxial particles with the distributed anisotropy barriers, magnetic moment $m$ of the sample will not
change in time $t$ exponentially, but according to
\begin{equation}
m(t) = m_{eq} + (m_{0}-m_{eq})\frac{\int_{0}^{\infty}\mathrm{d}Uf(U)U\e ^{-t/\tau }}{\int_{0}^{\infty}\mathrm{d}Uf(U)U} ,
\label{mt}
\end{equation}
where $m_{0}$ is the starting value of magnetic moment from which the
relaxation is measured and $m_{eq}$ is the equilibrium value toward which the
relaxation is going on. $\tau (U)$ is determined by Eq. \ref{tau}, where $U$ is the barrier height, whose distribution in the ensemble is given by $f(U)$.
One must have on mind that the applied magnetic field $H$ can distort the
barrier heights, but here this is neglected because the applied magnetic field of 0.03T is much lower than the anisotropy field of $\sim 0.5$T. At some specified temperature the exponential factor in Eq. \ref{mt} is essentially zero for $V<V_{B}$ and unity for $V>V_{B}$, where
\begin{equation}
V_{B} = \frac{kT}{K}\ln \left( \frac{t}{\tau _{0}} \right) .
\label{vblock}
\end{equation}
Hence, the magnetic moment in time changes according to:
\begin{equation}
m(t) = m_{0} - (m_{0}-m_{eq})
\frac{\int_{0}^{V_{B}}\mathrm{d}Vf(V)V}{\int_{0}^{\infty}\mathrm{d}Vf(V)V},
\label{logrelax}
\end{equation}
that is dependent on $\ln t$ when Eq. \ref{vblock} is inserted. The barrier height $U$ was translated to volume $V$ using the same reasoning as in calculation of the effective magnetic anisotropy density. Moreover, the result of Eq. \ref{logrelax} is the dependence of the magnetic moment $m$ on the combination $T\ln t$.

The other result of this reasoning is that if $f(V)V$ drops rapidly above
certain volume $V_{max}$, then the system can be characterized by blocking
temperature
\begin{equation}
T_{B} = \frac{K V_{max}}{k\ln (t/\tau _{0})}.
\label{tb}
\end{equation}
This is also a more exact confirmation of validity of the performed procedure of determination of the magnetic anisotropy density from measured $T_{B}$ using the top radius in the ensemble of particles.

For the description of the time-evolving of systems in the metastable states
with the distributed energy barriers, the so-called magnetic viscosity was
introduced (for example, see Ref. \cite{viscosity}), defined as:
\begin{equation}
S=-\frac{1}{m_{0}-m_{eq}}\frac{dm(t)}{d(\ln t)}.
\label{viscodef}
\end{equation}

This equation is customised to include properly the relaxation from the initial
measured state $m_{0}>0$ to the equilibrium state $m_{eq}<0$ of magnetic moment
of the sample. For $m_{eq}$ the value from the FC curve (measured in the same applied magnetic field as the relaxation) at the specific temperature is used. Although $m_{eq}$ does not change appreciably, it is included because of consistency. $m_{0}$ is the value of first measured point. It is not equal to the initial magnetic moment before the changing of field (which is equal to the FC value), but this is the value from which the relaxation at actual temperature is observable using our experimental device. Before the measurement starts, immediately after the change of applied field, the system overcomes very quickly to $m_{0}$ and after that the slow relaxation is measurable. Actually, this property provides the changing of the wanted measurable region of relaxation by controlling the temperature.

There are many systems which belong to the class of logarithmic relaxation.
Common assumption is that as the system is going toward the equilibrium, it
comes to the greater and greater potential barriers. In the case of our system,
as the time passes the greater and greater particles relax more significantly
toward the equilibrium during some characteristic time-scale of the observation.

\begin{figure}
\begin{center}
\includegraphics*[width=7.5cm]{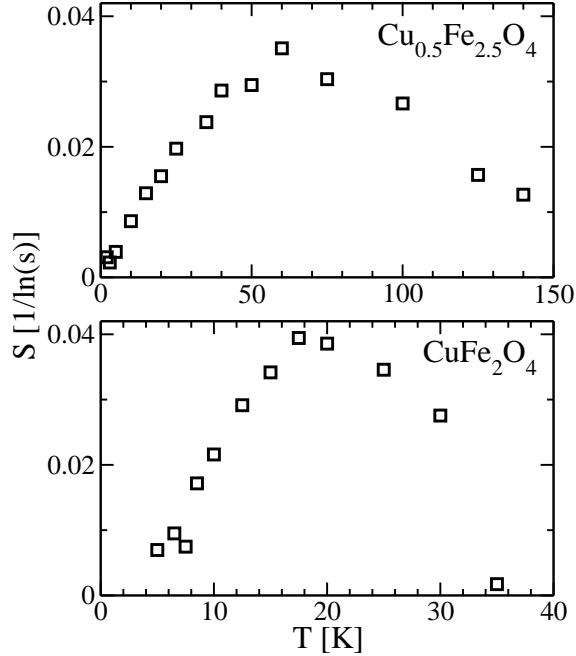}
\end{center}
\caption{The dependence of magnetic viscosity $S$ on temperature $T$ for Cu$_{0.5}$Fe$_{2.5}$O$_{4}$ and CuFe$_{2}$O$_{4}$.}
\label{visco}
\end{figure}

The dependence of magnetic viscosity $S$ on temperature $T$ is plotted in
Fig. \ref{visco}. For both samples, $S$ is proportional to $T$ below some
temperature. At certain temperature $S$ attains the maximum and above this
temperature it decreases quickly and drops to zero at $T_{B}$ of the sample. Above $T_{B}$ the system is in equilibrium from the point of view of the experimental set-up and $S=0$ holds. The maximum of $S$ corresponds to the temperature at which the most of relaxation happens during $\sim 3$ hours of measurement: the relaxation at this temperature is globally most effective.

Inserting Eq. \ref{vblock} and Eq. \ref{logrelax} into Eq. \ref{viscodef} and restricting to the low temperature region it follows (see Ref. \cite{mqtmm})
\begin{equation}
S = \frac{k T}{K\langle V \rangle } ,
\label{st}
\end{equation}
which predicts the linear dependence of $S$ on $T$. Here, $\langle V \rangle $
is the mean value of the volume of particles in ensemble. This result shows that the slope of $S(T)$ dependence is connected to the magnetic anisotropy. In order to obtain the effective $K$, linear regression is performed restricting to the points below the maximum of $S$. The intercepts are going near by origin within the standard error, and from the slope the $K\langle V \rangle /k$ is calculated: 1450K for Cu$_{0.5}$Fe$_{2.5}$O$_{4}$ and 357K for CuFe$_{2}$O$_{4}$. 

Taking into account just the uniform distribution
$g(r)$ over diameters of spherical particles from 5nm to 8nm, which is in accordance to measured diameters of (6.5$\pm $1.5)nm, the mean value of volume $\langle V \rangle = \int _{r_{min}}^{r_{max}}g(r)4\pi r^{2} \mathrm{d} r \simeq \frac{V_{max}}{4}$ is obtained, where $V_{max}=4\pi r_{max}^{3}/3$ in  calculation of $K$ from $T_{B}$ was used. The ratio $K V_{max}/K\langle V \rangle $, where $K V_{max}$ is calculated from the $T_{B}$ and $K\langle V \rangle $ is calculated from the linear fit of $S(T)$, is 3.67 for Cu$_{0.5}$Fe$_{2.5}$O$_{4}$ and 3.92 for CuFe$_{2}$O$_{4}$, which is
very close to 4. Slight deviation between these values comes from the very
roughly supposed distribution $g(r)$ over sizes of the particles.

Hence, both procedures give almost the same effective magnetic anisotropy
density $K$ within the standard error. This indicates that these two methods are mutually consistent. Also, this is another confirmation that $T_{B}$ is defined by $V_{max}$.

\begin{figure}
\begin{center}
\includegraphics*[width=7.5cm]{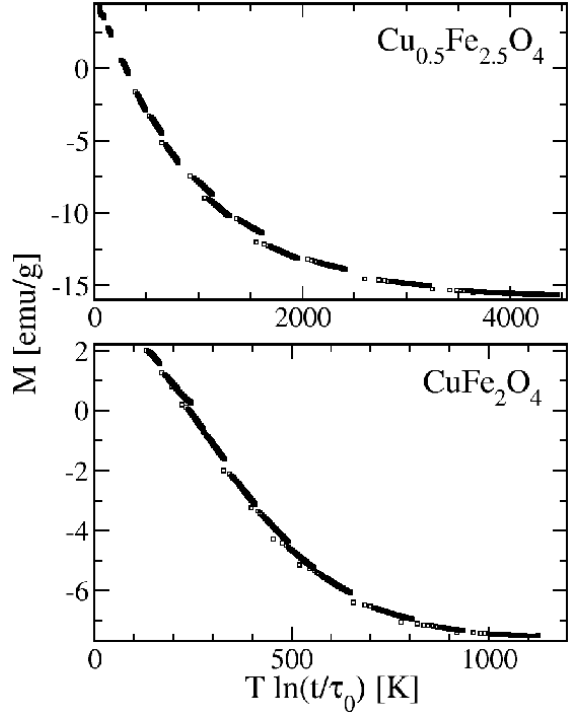}
\end{center}
\caption{Scaled relaxation curves at all measured temperatures for Cu$_{0.5}$Fe$_{2.5}$O$_{4}$ and CuFe$_{2}$O$_{4}$. Parameter $\tau _{0}=10^{-10}$s is used.}
\label{scaled}
\end{figure}

In the relaxation measurements both the temperature dependence and the time
dependence of magnetic moment are obtained. When the relaxation data for all
temperatures are plotted with the time-axis transformed as $T\ln (t/\tau_{0})$ using the appropriate temperature, all of the measured curves collapse onto a single scaling curve, as shown in Fig. \ref{scaled}. This result follows from thermal relaxation theory when Eq. \ref{logrelax} is considered. To obtain the collapse onto unique and continuous curve in the procedure of scaling the right $\tau _{0}$ must be chosen. Here, we used $\tau _{0}=10^{-10}$s, that is in accordance to many sources (see Ref. \cite{mqtmm} and the references therein).

Scaled plot allows to check whether the $T\ln (t/\tau_{0})$ scaling of the
relaxation, which is predicted by the thermal theory, holds at all temperatures. The missing of the scaling would point to the appearance of another mechanism of relaxation besides the thermal activation. This would indicate the presence of quantum tunneling of magnetic moment, as was seen at low temperatures in some experiments \cite{mqtexp1}. Also, quantum tunneling was used to explain the temperature independent magnetic viscosity $S$ below some temperature (0.1-5K) measured in the assemblies of nanoparticles \cite{mqtexp2}. According to rough estimation based on the arguments presented in Ref. \cite{tece}, the transition temperature for appearance of quantum relaxation regime in our samples is below 1K.

Finally, we observed the very slow variation of viscosity along the time during the measurement at constant temperature. Now, this can be understood from dynamical reasons looking at the scaled curve. Measurement at each temperature covers just a small part of the common curve, so that $M(\ln t)$ is quasi-linear during such a short time. However, the master curve in Fig. \ref{scaled} cannot reach beyond the equilibrium as long as we wait. Therefore, it must be curved. This is also observed as a very slight change of the slope of $M(\ln t)$ curves in Fig. \ref{relax} during the measurement at some of temperatures. Scaled curve actually covers a very wide time interval of several orders of magnitude compared to the measurement at only one temperature, therefore it is able to uncover this slow and slight change.

\section{Conclusion}
\label{conclusion}

All of measurements performed on the collection of nanoparticles of
Cu$_{0.5}$Fe$_{2.5}$O$_{4}$ and CuFe$_{2}$O$_{4}$ point to their superparamagnetic behaviour. It is expressed through mutual consistency of
the M\"{o}ssbauer spectra, FC-ZFC splitting, hysteresis and relaxation measurements. Qualitatively the same phenomena in both samples were observed, which are characteristic for the assemblies of magnetic nanoparticles.

However, quantitatively the difference in properties between two investigated samples is observed. It manifests in the different anisotropy and different
value of the saturation magnetization. Much greater anisotropy in Cu$_{0.5}$Fe$_{2.5}$O$_{4}$ than in CuFe$_{2}$O$_{4}$ is a consequence of the change of occupation of places inside A and B superlattices of the ferrimagnetic spinel structure with the copper and iron cations. The possible explanation for change of properties is the Jahn-Teller distortion in crystal lattice. Most probably in the nanosized particle under the action of the Laplace pressure, antiferromagnetic superstructure arises in the spherical particle, where the forces of exchange interactions are not sufficient to maintain the spontaneous magnetic moment ordering that is characteristic for fully built crystal structure of an edged particle. These zones affect the magnetic moments density and depend on the temperature.

The relaxation data at many different temperatures show the unique time
dependence of magnetic moment when appropriate scaling is performed. This is a
confident criterium for thermal relaxation over the distributed potential
barriers according to the described model. Moreover, it allows to follow what
happens with magnetic moment of the sample during very long times, covering
several orders of magnitude.

Exhibited mutual consistency of all performed different types of measurements is important point in the characterisation of new type of materials. Also,
presented phenomenological description of the magnetic relaxation provides an
useful link between the experimental results and microscopic model.

Future work will search for the more precise answer to questions opened here. It will research in more details the hysteretic phenomena, blocking at higher
applied magnetic fields and extension of the effective anisotropy with more
details.

\appendix

\section*{Acknowledgments}
\label{ackn}
This work has been supported by the Croatian Ministry of Science and Technology
through project 0119258, Belgian Fund for Scientific Research FWO and Bulgarian National Foundation of Scientific Research Grant TH-01/01/2003.


\begin{thebibliography}{00}


\bibitem{interscaling}
T. Jonsson, P. Svendlindh, M.F. Hansen, Phys. Rev. Lett. {\bf 81} (1998) 3976

\bibitem{interallcomp}
R.N. Panda, N.S. Gajbhiye, G. Balaji, J. Alloy. Compd. {\bf 326} (2001) 50

\bibitem{interdipexch}
D. Kechrakos, K.N. Trohidou, J. Magn. Magn. Mater. {\bf 262} (2003) 107

\bibitem{glassprl}
B. Mart\'{i}nez, X. Obradors, Ll. Balcells, A. Rouanet, C. Monty, Phys. Rev. Lett. {\bf 80} (1998) 181

\bibitem{glassprb}
M. Ulrich, J. Garc\'{i}a-Otero, J. Rivas, A. Bunde, Phys. Rev. B {\bf 67} (2003) 024416

\bibitem{mechan}
S.J. Stewart, M.J. Tueros, G. Cernicchiaro, R.B. Scorzelli, Solid State Commun. {\bf 129} (2004) 347

\bibitem{memory}
O. Cador, F. Grasset, H. Haneda, J. Etourneau, J. Magn. Magn. Mater. {\bf 268} (2004) 232

\bibitem{memoerasing}
M. G. del Muro, X. Batlle, A. Labarta, Phys. Rev B {\bf 59} (1999) 13584

\bibitem{collective}
S.V. Gorobets, I.A. Melnichuk, J. Magn. Magn. Mater. {\bf 182} (1998) 61

\bibitem{qtmprl}
E.M.Chudnovsky, L. Gunther, Phys. Rev. Lett. {\bf 60} (1988) 661

\bibitem{qtm94}
Quantum tunneling of magnetization - QTM'94, NATO Science Series: E 301, Ed. L. Gunther, B. Barbara, Proceedings of the NATO Advanced Research Workshop, Grenoble and Chichilianne, France, June 27-July 2, 1994

\bibitem{japcufe2o4}
G.F. Goya, H.R. Rechenberg, J. Appl. Phys. {\bf 84} (1998) 1101

\bibitem{nanostructcufe2o4}
G.F. Goya, H.R. Rechenberg, Nanostruct. Mater. {\bf 10} (1998) 1001

\bibitem{lanlif}
L.D. Landau, E.M. Lifschitz, Electrodynamics of continuous media, Butterworth-Heinenann, Oxford, U.K., 1995

\bibitem{singledomain}
C. Kittel, Rev. Mod. Phys. {\bf 21} (1949) 541

\bibitem{fluctdomain}
W.F. Brown, Phys. Rev. {\bf 130} (1963) 1677

\bibitem{supermater}
I, Nedkov, Supermaterials, Kluwer Acad. Publ. (eds. R. Cloots et al.), NATO
Science Serie II Mathematics, Physics and Chemistry (2000) vol.{\bf 8}, 115

\bibitem{activvol}
A.M. de Witte, K. O'Grady, G.N. Coverdale, R.W. Chantrell, J. Magn. Magn. Mater. {\bf 88} (1990) 183

\bibitem{jpcm}
J.Z. Jiang, G.F. Goya, H.R. Rechenberg, J. Phys.:Codens. Matter {\bf 11} (1999)
4063

\bibitem{sizeeffect}
A.-F. Lehlooh, S.H. Mahmood, J.M. Williams, Physica B {\bf 321} (2002) 159

\bibitem{dynamichyster}
C. Caizer, I. Hrianca, Eur. Phys. J. B {\bf 31} (2003) 391

\bibitem{cellmodel}
E. Gorter, Adv. Phys. {\bf 6} (1957) 336

\bibitem{mqtmm}
E.M. Chudnovsky, J. Tejada, Macroscopic Quantum Tunneling of the Magnetic Moment, Cambridge University Press, Cambridge, U.K., 1998, pp. 89-94

\bibitem{viscosity}
J. Tejada, X.X. Zhang, E.M. Chudnovsky, Phys Rev. B {\bf 47} (1993) 14977

\bibitem{mqtexp1}
X.X. Zhang, R. Ziolo, E.C. Kroll, X. Bohigas, J. Tejada, J. Magn. Magn. Mater {\bf 140-144} (1995) 1853

\bibitem{mqtexp2}
J. Tejada, X.X. Zhang, J. Magn. Magn. Mater {\bf 140-144} (1995) 1815

\bibitem{tece} 
B. Barbara, E.M. Chudnovsky, Phys. Lett. A, {\bf 145} (1990) 205







\end{thebibliography}
\end{document}